\documentclass[journal=ancac3,layout=twocolumn,manuscript=article]{achemso}

\usepackage{amsmath}
\usepackage{braket}
\usepackage{txfonts}
\usepackage{bm}
\usepackage{color}
\usepackage{verbatim}
\usepackage[font=small]{caption}
\definecolor{myblue}{rgb}{0,0,1}
\usepackage[breaklinks=true,colorlinks=true,linkcolor=myblue,urlcolor=myblue,citecolor=myblue]{hyperref}

\title{Surface Hopping Simulations Show Valley Depolarization Driven by Exciton--Phonon Resonance}

\author{Alex Krotz}
\affiliation{Department of Chemistry, Northwestern University, 2145 Sheridan Road, Evanston, Illinois 60208, USA}

\author{Roel Tempelaar}
\affiliation{Department of Chemistry, Northwestern University, 2145 Sheridan Road, Evanston, Illinois 60208, USA}
\email{roel.tempelaar@northwestern.edu}

\makeatletter
\let\l@addto@macro\relax
\makeatother
\usepackage[fontsize=11pt]{scrextend}

\begin{document}

\maketitle

\begin{abstract}
Resonances between excitonic transitions and nuclear coordinates have been shown to drive a variety of excited-state dynamical phenomena in molecular systems. Here, we report mixed quantum--classical simulations showing similar resonances to primarily contribute to valley depolarization in monolayer MoS$_2$. The applied simulation framework combines reciprocal-space surface hopping with microscopic models of the quasiparticle band structure, electron--hole interactions, and carrier--phonon interactions, parametrized against \emph{ab initio} calculations. This enables low-cost excited-state dynamics simulations that are microscopic, non-Markovian, and non-perturbative in the carrier--phonon interaction. The framework furthermore retains explicit information on transient phonon occupancies, through which we show a resonance between the dominant optical phonon branch and the lowest exciton band to largely drive valley depolarization, by activating a Maialle--Silva--Sham mechanism. Resulting valley polarization times are consistent with experimental measurements across temperatures.
\end{abstract}

\section{Introduction}

Excited-state dynamical phenomena in molecules and materials hold myriad technological opportunities, but the underlying interplay of electronic and nuclear coordinates continues to challenge our fundamental understanding. Interactions between electronic carriers and nuclear modes can often be considered weak, in which case phenomena are well-described by a variety of numerical methods that minimize computational cost by perturbative and/or Markovian approximations \cite{breuer2002theory}. However, such approximations need to be relaxed once interactions are strong, or when electronic and nuclear coordinates become resonant \cite{may2023charge}. In recent years, resonances have been reported to drive a variety of excited-state phenomena in molecular systems. Examples are to be found in photosynthetic complexes, where energy transfer and optical responses have been shown to be governed by high-frequency intramolecular vibrations whose quanta are resonant with energy gaps between electronic transitions \cite{womick2011vibronic, christensson2012origin, tiwari2013electronic, fuller2014vibronic, romero2014quantum, palecek2017quantum, thyrhaug2018identification}. A similar ``vibronic resonance'' has been shown to promote the multiplication of excitons (bound electron--hole pairs) in organic molecular crystals \cite{bakulin2016real, fujihashi2017effect, morrison2017evidence, tempelaar2018vibronic, unger2022modulating}, which may enable photovoltaic cells to overcome the Shockley--Queisser limit \cite{hanna2006solar}. In this Article, we expose analogous resonance effects in the realm of crystalline solids, by an \emph{in silico} demonstration of exciton--phonon resonances as a primary driver of valley depolarization in monolayer MoS$_2$.

\begin{figure*}
    \centering
    \includegraphics{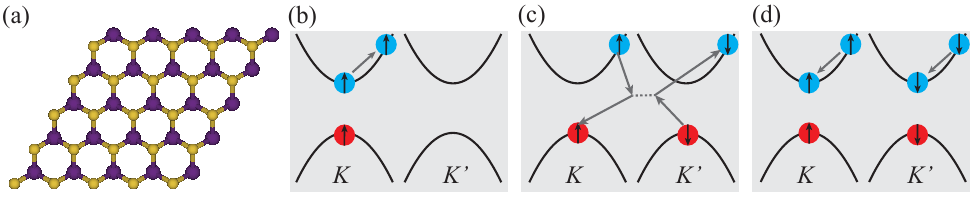}
    \caption{Top-down view of the atomic structure of monolayer MoS$_2$ with transition metals and chalcogens shown in yellow and purple, respectively (a). Also shown is a cartoon depiction of the Maialle--Silva--Sham (MSS) mechanism for valley depolarization, including the valley-resolved lowest and highest conduction and valence spin-bands, respectively, with electrons (holes) shown in blue (red), and with spins indicated. The MSS mechanism involves two phonon-induced electron (or hole) scattering events (b and d) combined with electron--hole exchange interactions (c), distributing a valley-polarized initial excitation across both valleys.}
    \label{fig:FIG_MSS}
\end{figure*}

MoS$_2$ is a member of the family of transition-metal dichalcogenides (TMDs), which in their monolayer form \cite{novoselov2005two} represent direct-bandgap semiconducting analogs of graphene \cite{mak2010AtomicallyThinMoS, splendiani2010emerging}. The associated two-dimensional lattice assumes a hexagonal shape, with sites alternating between transition metals and pairs of chalcogens, as depicted in Fig.~\ref{fig:FIG_MSS} (a). The bandgaps are located at the corners of the hexagonal Brillouin zone \cite{li2007electronic, lebegue2009electronic}, which due to inversion-symmetry breaking form inequivalent pairs \cite{xiao2012coupled} labeled $K$ and $K'$. The band regions around the $K$ and $K'$ points are referred to as ``valleys'' \cite{yao2008valley}, and are subject to an opposite topology that promotes selectivity to right- and left-handed circularly-polarized optical excitation, respectively \cite{yao2008valley}. Additionally, sizable spin-orbit coupling \cite{zhu2011giant} induces opposite spin-splitting of the quasiparticle bands around $K$ and $K'$, yielding an interlocking of spin and valley degrees of freedom \cite{xiao2012coupled}. Combined with a low dielectric screening environment enabled by the material's low dimensionality \cite{chernikovExcitonBindingEnergy2014, latini2015excitons}, these effects allow for the selective creation of stable excitons \cite{chernikovExcitonBindingEnergy2014} with well-defined spin- and valley-polarization through the use of circularly-polarized light \cite{xiao2012coupled, cao2012valley, kioseoglou2012valley, mak2012control, sallen2012robust, zeng2012valley}. Moreover, the interlocking of spin and valley degrees of freedom renders such excitons stable against depolarization \cite{xiao2012coupled}, at least in principle, offering exciting prospects for spintronic and valleytronic applications.

While TMDs present the principles necessary to conserve valley polarizations, experimental measurements have shown depolarization to occur rapidly and effectively \cite{zeng2012valley, kioseoglou2012valley, kumar2014valley, zhu2014exciton, mai2014many, mai2014exciton, dalconte2015ultrafast, yan2015valley, wang2015polarization, schmidt2016ultrafast, pflechinger2017valley, kim2017valley, wang2018intravalley, mccormick2018imaging, tornatzky2018resonance}. The dominant depolarization contribution is commonly ascribed \cite{yu2014valley, selig2020suppresion} to a Maialle--Silva--Sham (MSS) mechanism \cite{maialle1993exciton}, driven by intervalley exchange interactions. Such interactions vanish for excitons with small center-of-mass (COM) wavevector (lattice momentum), including those that result from optical excitation, yet scattering with phonons induces a wavevector change that render them effective. Accordingly, valley depolarization ensues through a cascaded pathway that depends sensitively on carrier--phonon interactions, as depicted in Fig.~\ref{fig:FIG_MSS} (b-d). Previous studies have theoretically and computationally evaluated valley depolarization times in TMDs with incorporation of phonons \cite{molina2017ab, selig2020suppresion, yang2020exciton, chen2022first}, showing favorable agreement with temperature-dependent experimental results, yet much remains to be learned about the exact mechanisms through which phonons steer this process. Notably, recent measurements on MoS$_2$ have shown carrier--phonon interactions to be sizable \cite{li2021exciton}, underscoring a specific need for non-perturbative and non-Markovian simulation studies.

Our approach to modeling valley depolarization in MoS$_2$ is to retain a nonperturbative and non-Markovian treatment of carrier--phonon interactions, while minimizing the computational cost by invoking the classical approximation for the phonon coordinates. In doing so, we apply a recently-developed mixed quantum--classical (QC) simulation framework that offers microscopic detail for both carrier and phonon coordinates \cite{krotz2024mixed}. This affords a wavevector-resolved analysis of the phononic occupancies, through which we discover prominent beatings attributed to non-Markovianity, accompanied by a near-monotonic growth of features coinciding with exciton--phonon resonances. Such resonances can be included or excluded by means of truncations of the phononic Brillouin zone, which provides a means to assess their effect on valley depolarization. Through this approach, we find valley depolarization in MoS$_2$ to be predominantly driven by a resonance between the lowest dispersed exciton branch and the dominant nondispersed optical phonon branch.

\section{Theoretical framework}

The applied QC framework was detailed exhaustively in a previous work by the authors \cite{krotz2024mixed}, where a study was presented on the temperature-dependent optical linewidths of MoS$_2$. We will proceed with a quick summary of this framework, while elaborating on any modifications made in order to facilitate the evaluation of valley depolarization in the same material. The present iteration of this framework is implemented as a Python script that uses the QC Lab package \cite{krotzQCLabPython2026}. This script serves as the platform for all reported calculations, and is made publicly available \cite{repo2026}. We also note that the applied framework is broadly reflective of a recent concerted effort to apply QC dynamics to crystalline solids \cite{nie2014ultrafast, prezhdo2021modeling, krotz2021reciprocal, xie2022surface, lively2024revealing, chen2024floquet}.

The QC framework describes the phonons classically, while reserving a quantum treatment for the electronic coordinates. A reciprocal-space formulation of the resultant phononic and electronic equations \cite{krotz2021reciprocal, krotz2022reciprocal} then allows dynamics to be evaluated at radically-reduced computational cost by means of truncations of the Brillouin zone \cite{qiu2016screening, tempelaar2019many-body, krotz2021reciprocal, krotz2022reciprocal}. Accordingly, free motion of the phonons is governed by the Hamiltonian function
\begin{equation}
    H_{\mathrm{ph}}(\bm{z}) = \sum_{\bm{k},\mu}\omega_{\bm{k},\mu}z^{*}_{\bm{k},\mu}z_{\bm{k},\mu},
\end{equation}
where $z_{\bm{k},\mu}$ is a complex-valued classical coordinate \cite{krotz2021reciprocal, krotz2022reciprocal, miyazaki2024unitary} for the phonon in branch $\mu$ and at wavevector $\bm{k}$.

In the absence of phonons, excitonic eigenstates are expanded in terms of electron and hole excitations of the Fermi vacuum as
\begin{equation}
    \ket{\Phi_n} = \sum_{\bm{k},c,v} A_{\bm{k},c,v}^n \hat{c}^{\dagger}_{\bm{k} + \bm{k}_n,c} \hat{c}_{\bm{k},v}\ket{0},\label{eq:exciton}
\end{equation}
with $\hat{c}^{\dagger}_{\bm k,c(v)}$ and $\hat{c}_{\bm k,c(v)}$ representing the operators for creation and annihilation, respectively, of an electron in spin-band $c(v)$ with wavevector $\bm k$. In Eq.~\ref{eq:exciton}, $n$ labels the excitonic state and $\bm{k}_n$ represents the associated total (i.e., COM) wavevector. Within the basis of excitonic eigenstates, the excitonic Hamiltonian operator is given by
\begin{equation}
    \hat{H}_{\mathrm{x}} = \sum_n E_n \hat{C}^{\dagger}_n\hat{C}_n,
\end{equation}
where the excitonic creation operator follows from $\hat{c}^{\dagger}_{\bm{k}+\bm{\kappa},c}\hat{c}_{\bm{k},v}=\sum_{n}A^{n*}_{\bm{k},c,v}\hat{C}^{\dagger}_{n}\;\delta_{\bm{k}_n,\bm{\kappa}}$.\footnote{We note that the Kronecker delta was inadvertently omitted in Ref.~\citenum{krotz2024mixed}.} The Hamiltonian operator governing the exciton--phonon interaction is given by
\begin{equation}
    \hat{H}_{\mathrm{x-ph}} = \sum_{n,m,\mu}G_{n,m,\mu} \hat{C}^{\dagger}_n \hat{C}^{\phantom{\dagger}}_m \left(z^{*}_{-\bm{k}_{nm},\mu} + z_{\bm{k}_{nm},\mu}\right),
    \label{eq:H_x_ph}
\end{equation}
where $k_{nm} = k_n - k_m$ and with
\begin{align}
    G_{n,m,\mu} \equiv & \sum_{\bm{k},c,v} \Big[g_{\bm{k}+\bm{k}_{m},\bm{k}_{nm},\mu}^{c} \left(A^n_{\bm{k},c,v}\right)^* A^m_{\bm{k},c,v} \label{eq:Ex-ph} \\
    &- g^{v}_{\bm{k}-\bm{k}_{n},\bm{k}_{nm},\mu} \left(A^n_{\bm{k}-\bm{k}_{n},c,v}\right)^* A^m_{\bm{k}-\bm{k}_{m},c,v} \Big].\nonumber
\end{align}
Here, $g^{c(v)}_{\bm{k},\bm{\kappa},\mu}$ is the spin-conserving intraband matrix element for the scattering of an electron in the conduction (valence) band onto a phonon in branch $\mu$ with wavevector $\bm{\kappa}$, thereby undergoing a change of wavevector from $\bm{k}$ to $\bm{k}+\bm{\kappa}$.

In our previous work \cite{krotz2024mixed}, the QC framework was implemented by means of Ehrenfest's theorem \cite{ehrenfest1927bemerkung, tully1998MixedQuantumClassical}, yielding a mean-field approach to QC dynamics, which is expected to yield sufficient accuracy at the ultrashort timescales relevant to optical line broadening. However, valley depolarization dynamics proceeds at a timescale where inaccuracies of mean-field dynamics become apparent \cite{parandekar2005mixed, parandekar2006detailed, krotz2022reciprocal}. For the present study, we have therefore extended our framework by incorporating the fewest-switches surface hopping (FSSH) approach \cite{tully1990molecular} to QC dynamics. This approach relies on the adiabatic propagation of a quantum state, interrupted by stochastic hops of this state among adiabats in order to account for nonadiabatic effects \cite{tully1990molecular, hammes1994proton}. We have recently generalized the FSSH approach in terms of the complex-valued classical coordinates necessary for the reciprocal-space formulation of QC dynamics, details of which can be found in Refs.~\citenum{krotz2022reciprocal} and \citenum{miyazaki2024unitary}.

Importantly, the electron and hole Bloch vectors in MoS$_2$ exhibit geometric phase effects, which render the total Hamiltonian operator, $\hat{H}_\mathrm{x}+\hat{H}_\mathrm{x-ph}(\bm{z})$, inherently complex-valued. Our previous mean-field implementation of the QC framework \cite{krotz2024mixed} is immune to complex-valued Hamiltonian operators. By contrast, however, special care needs to be taken under the currently-applied FSSH approach. FSSH relies on a renormalization of classical coordinates upon a hop between adiabats, with rescaling directions determined based on derivative couplings. In the presence of geometric phase effects, the necessity of mapping complex-valued derivative couplings to real-valued vectors is prone to introduce a gauge ambiguity, which has been the topic of study \cite{miao2019extension, bian2021modeling, bian2022incorporating, krotz2024TreatingGeometricPhase}. Our QC framework builds on our previous work \cite{krotz2024TreatingGeometricPhase}, establishing a protocol for fixing the gauge based on decomposing Hamiltonians into generators of the special unitary group (at a dimensionality of the corresponding Hilbert space), and by minimizing $\bm{z}$ dependent fluctuations associated with the imaginary generators. Notably, for the case of MoS$_2$, this minimization can be performed trivially, so long as each quasiparticle band is represented in terms of both spin replicas and within both the  $K$ and $K'$ valleys, while also ensuring that the intervalley exchange interaction is included in the Hamiltonian.

The incorporation of the FSSH approach within our QC framework delivers a versatile platform for modeling the dynamics of materials at a favorable accuracy-to-cost ratio. An interfacing of this framework with a full \emph{ab initio} treatment of a material is straightforward in principle. While we plan to explore such interfacing in future studies, we minimize the complexity of the current MoS$_2$ implementation by employing models for the quasiparticle band structure, electron--hole interaction elements, and carrier--phonon interaction elements. As such, we closely follow the approach that was comprehensively described in our previous study on the optical linewidths of MoS$_2$ \cite{krotz2024mixed}. In short, the conduction and valence bands are obtained based on a spin-generalized massive Dirac-like Hamiltonian parametrized against \emph{ab initio} calculated band structures \cite{xiao2012coupled, ochoa2013spin}. This minimal representation serves to accurately represent the $K$ and $K'$ valley regions that govern the rapid valley depolarization dynamics of MoS$_2$ \cite{selig2020suppresion}. The electron--hole interaction elements are taken in the Rytova--Keldysh potential form \cite{rytova1967screened, keldysh1979coulomb}, similarly parametrized against \emph{ab initio} calculations \cite{berkelbach2013theory}.

As before \cite{krotz2024mixed}, carrier--phonon interactions are described within the deformation potential approximation. Accordingly, the matrix element associated with the conduction (valence) band is given by
\begin{equation}
    g^{c(v)}_{\bm{k},\bm{\kappa},\mu} = \sqrt{\frac{\hbar}{2mN\omega_{\kappa,\mu}}}\left(D_{0,\mu}^{c(v)} + D_{1,\mu}^{c(v)}\kappa\right)\exp\left(i\phi^{c(v)}_{\bm{k},\bm{\kappa}}\right).
    \label{eq:g_quasiparticle}
\end{equation}
Here and henceforth, we denote vector norms in regular font, e.g., $\kappa=\vert\bm{\kappa}\vert$. Furthermore, in Eq.~\ref{eq:g_quasiparticle}, $\phi^{c(v)}_{\bm{k},\bm{\kappa}}$ accounts for the global gauge associated with the electron (hole) eigenstate \cite{krotz2024mixed}, and $D^{c(v)}_{0,\mu}$ and $D^{c(v)}_{1,\mu}$ are the zeroth- and first-order deformation potential constants, respectively. In a recent computational study on monolayer MoSe$_2$, both acoustic and optical phonon modes were incorporated, yielding an approximate linear dependence of optical linewidths with temperature, from which it was inferred that optical phonons contribute only weakly to the observed line broadening \cite{selig2016ExcitonicLinewidthCoherence}. Based on the structural similarity between MoSe$_2$ and MoS$_2$, optical phonons were therefore excluded from the QC framework applied in our previous study \cite{krotz2024mixed}, while the relevant transverse acoustic (TA) and longitudinal acoustic phonons (LA) were incorporated as a single branch represented as a Debye mode, following a precedent in the literature \cite{li2013IntrinsicElectricalTransporta}. Accordingly, the mode energy was governed by $\omega_{\bm{k},\mathrm{ac}}= v k$, with $v$ as the sound velocity, and the zeroth-order deformation potential constants were omitted. In order to holistically investigate the phonon effect on valley depolarization, in the present study we expand the phononic representation by the addition of a single optical branch. This branch is taken to be representative of relevant transverse optical (TO), longitudinal optical (LO), and homopolar ($A_1$) phonons, again following precedent in the literature \cite{li2013IntrinsicElectricalTransporta}, and is described as an Einstein mode. Accordingly, the mode energy is taken to be wavevector-independent, $\omega_{\bm{k},\mathrm{op}}=\omega_0$, and the first-order deformation potential constants are omitted.

\begin{table}
\centering
\begin{tabular}{ |c|c|c| } 
 \hline
 Parameter &  Value & Ref.\\ 
 \hline 
 $v$ & $66$ \AA~ps$^{-1}$ & ~\citenum{li2013IntrinsicElectricalTransporta}~ \\ 
 $\omega_0$ & $48.0$ meV & ~\citenum{li2013IntrinsicElectricalTransporta}~ \\ 
 $m$ & $2.66 \cdot 10^{-25}$ kg & -- \\
 $D_{1,\mathrm{ac}}^{c}$ & $3.18$ eV & ~\citenum{jinIntrinsicTransportProperties2014}~ \\ 
 $D_{1,\mathrm{ac}}^{v}$ & $-1.77$ eV & ~\citenum{jinIntrinsicTransportProperties2014}~ \\
 $D_{0,\mathrm{op}}^{c}$ & $5.74$ eV \AA$^{-1}$ & ~\citenum{jinIntrinsicTransportProperties2014}~ \\ 
 $D_{0,\mathrm{op}}^{v}$ & $-3.76$ eV \AA$^{-1}$ & ~\citenum{jinIntrinsicTransportProperties2014}~ \\ 
 \hline
\end{tabular}
\caption{Phonon parametrization applied in our calculations, including the references where parameters were taken from. The unit cell mass $m$ was taken to be the total mass of one molybdenum atom and two sulphur atoms.}
\label{tab:parms}
\end{table}

In modeling MoS$_2$ using our QC framework, the applied parameters are adopted directly from our previous work \cite{krotz2024mixed}, with the exception of those pertaining to the phonons, which we summarize in Tab.~\ref{tab:parms}. Here, we note that the deformation potential constants taken for the acoustic phonon branch differ from those applied previously by a factor of $\sqrt{2}$. In inferring the values based on Refs.~\citenum{li2013IntrinsicElectricalTransporta} and \citenum{jinIntrinsicTransportProperties2014}, we previously misinterpreted these constants as being mode-resolved with respect to the TA and LA phonons, whereas in actuality they represent the aggregate values to be adopted for the single acoustic branch, as done here \footnote{A correction to Ref.~\citenum{krotz2024mixed} is in preparation.}. We further note that, as before \cite{krotz2024mixed}, these deformation potential constants have been divided by $\sqrt{2}$ (compared to Refs.~\citenum{li2013IntrinsicElectricalTransporta} and \citenum{jinIntrinsicTransportProperties2014}). Such is done as a means to omit the piezoelectric contribution \cite{lengersTheoryAbsorptionLine2020}, which is automatically included in the electron and hole deformation potential constants, but which does not contribute for excitons due to the opposite charges of the constituent electron and hole.

The parametrization of the optical phonon branch is similarly based on values reported in Refs.~\citenum{li2013IntrinsicElectricalTransporta} and \citenum{jinIntrinsicTransportProperties2014}. In doing so, the deformation potential constants are specifically based on the values associated with intravalley electron and hole scattering pathways, seeing that these pathways are expected to dominate in the MSS mechanism (cf.~Fig.~\ref{fig:FIG_MSS}). For the associated mode energy, on the other hand, an average is taken over the different Brillouin zone regions (where it should be noted that energy fluctuations are minor across these regions). While the optical phonon branch is representative of the TO, LO, and $A_1$ phonon modes, the LO mode couples mainly via a Fr\"ohlich-type interaction, which experiences a destructive interference for excitons \cite{kaasbjerg2013AcousticPhononLimited}, much like the aforementioned piezoelectric contribution to the acoustic branch. Similarly to that case, this Fr\"ohlich contribution is automatically included in the electron and hole deformation potential constants reported in Refs.~\citenum{li2013IntrinsicElectricalTransporta} and \citenum{jinIntrinsicTransportProperties2014}, but can be divided out straightforwardly. Given that the LO mode contributes approximately one third of the coupling magnitude in Ref.~\citenum{kaasbjergPhononlimitedMobilityType2012}, a division by $\sqrt{3/2}$ is performed.

In our calculations, the two-dimensional Brillouin zone is sampled at a resolution of $N_\mathrm{k} \times N_\mathrm{k}$. A key feature of the applied QC framework is that it allows the basis of quantum states and classical coordinates to be efficiently truncated \cite{krotz2024mixed}. In modeling the fast component of valley depolarization in MoS$_2$, we apply a Brillouin zone truncation radius for electrons and holes around the $K$ and $K'$ points, denoted $R_\mathrm{k}$ \cite{krotz2024mixed}, rationalized by the principle that relevant excitonic states involve quasiparticles localized at those band extrema. In addition, we apply a similar truncation radius to the COM wavevector of the exciton, denoted $R_\mathrm{x}$ \cite{krotz2024mixed}. Since the effect of phonons is to receive or donate lattice momentum upon electronic transitions, the phonon coordinates are then automatically truncated by omitting all coordinates whose wavevector do not match a lattice momentum difference between any of the remaining electron or hole state combinations. Altogether, this equips us with three parameters with respect to which convergence is to be ensured.

\section{Results}

Before turning our attention to valley depolarization, we first revisit the optical linewidths of MoS$_2$, focusing on the principal peak dominating the low-energy absorption spectrum. Compared to our previous investigation of this peak \cite{krotz2024mixed}, the modeling framework now differs in various aspects. First, the mean-field approach taken previously is now replaced by FSSH. Second, the phonon representation has been adjusted, and extended to include the optical phonon branch. The associated energy, $\omega_0$, significantly exceeds that of typical acoustic energies, and surpasses values of the thermal quantum especially at the lower range of relevant temperatures, as a result of which vacuum fluctuations become significant. Consequently, whereas we previously applied the Boltzmann distribution to sample initial classical coordinates, we now resort to the Wigner quasiprobability distribution, which incorporates vacuum fluctuations, and which is given by
\begin{equation}
    P(\bm{z})\propto\prod_{\bm{k},\mu}\exp\left[ -2\tanh\left(\frac{\omega_{\bm{k},\mu}}{2k_{\mathrm{B}}T}\right)z^{*}_{\bm{k},\mu}z_{\bm{k},\mu}\right].
\end{equation}
Here, $k_\mathrm{B}$ and $T$ represent the Boltzmann constant and the temperature, respectively.

Absorption spectra are obtained through a Fourier transform of the electronic momentum correlation function (Supporting Information and Ref.~\citenum{krotz2024mixed}), with the latter being calculated over an interval spanning 2.5 ps using a timestep of 0.025 fs. In doing so, an average is taken over 10\;000 QC trajectories. The resulting spectrum is convolved with a Lorentzian in order to account for non-phononic contributions, using a temperature-independent full width at half maximum (FWHM) of 40~meV. The total temperature-dependent linewidth is  then determined by taking the FWHM of a Lorentzian fit to the principal absorption peak. As done in our previous study \cite{krotz2024mixed}, we determine linewidths in the asymptotic limit of convergence parameter values, $\gamma_\infty$, by conducting calculations at various finite parameter values followed by an extrapolation to infinity based on exponential fittings (Supporting Information).

\begin{figure}[t!]
    \centering
    \includegraphics{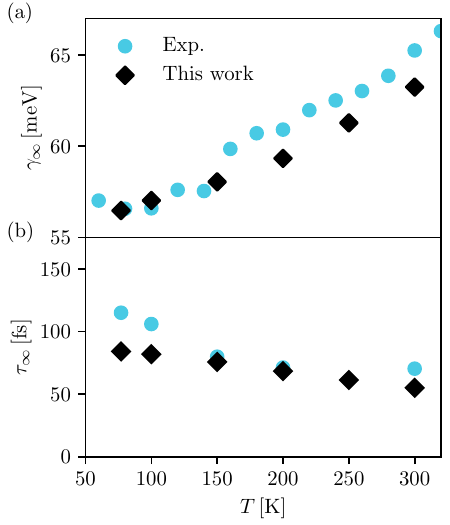}
    \caption{Asymptotic optical linewidths (a) and valley depolarization times (b) obtained through a fitting to calculations at varying values of the truncation parameters (black diamonds). Experimentally-measured values \cite{dey2016OpticalCoherenceAtomicMonolayer, dalconte2015ultrafast} are also shown (cyan circles).}
    \label{fig:FIG_LW_TRFR}
\end{figure}

Shown in Fig.~\ref{fig:FIG_LW_TRFR} (a) are calculated values of $\gamma_\infty$ against experimentally-measured linewidths \cite{dey2016OpticalCoherenceAtomicMonolayer} for temperatures ranging from 77~K to 300~K. Agreement between calculations and measurements is seen to be favorable across this temperature range. Notably, the calculations largely reproduce the linear increase of the measured linewidths with temperature. While this may appear to confirm that temperature-dependent line broadening is dominated by acoustic phonons, we instead find optical phonons to provide a comparable contribution (Supporting Information). Such is rationalized by the optical phonon vacuum fluctuations contributing significantly to the linewidths at low temperatures, resulting in line broadening behaviors that differ markedly from those expected at elevated temperatures.

We now proceed to consider valley polarization dynamics. In order to establish the initial electronic quantum state, we first recognize that the two lowest-energy excitonic eigenstates ($\Phi_n$) form a degenerate pair. The states in this pair each have a zero COM wavevector, but they differ in terms of their valley character, with one being localized around $K$ and the other around $K'$. As such, they correspond to the A excitons that are selectively addressable by a circularly-polarized excitation pulse \cite{xiao2012coupled}. As an initial condition, we take the A exciton associated with the $K$ valley. Subsequently, we evaluate dynamics by means of FSSH, during which the valley polarization is transiently monitored by considering the difference in occupancies between the A excitons localized in the $K$ and $K'$ valleys, $\rho_{\mathrm{A},K}-\rho_{\mathrm{A},K'}$. Probing this quantity in combination with the applied initial condition is representative of a transient valley-polarization measurement by means of time-resolved Faraday rotation (TRFR).

Asymptotic valley depolarization times, $\tau_\infty$, are obtained through an extrapolation analysis, analogous to that applied for $\gamma_\infty$ (Supporting Information). Results are shown in Fig.~\ref{fig:FIG_LW_TRFR} (b) against the rapid and dominant valley depolarization time component measured through TRFR experiments \cite{dalconte2015ultrafast} across temperatures ranging from 77 K to 300 K. The level of agreement between calculations and measurements is again seen to be favorable, and comparable to that reached for the optical linewidths. Notably, the calculations capture the weak yet appreciable drop of the depolarization time with increasing temperature.

\begin{figure}[t!]
    \centering
    \includegraphics{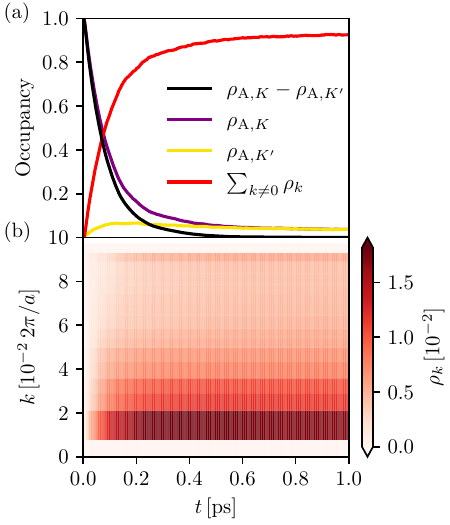}
    \caption{Calculated transient exciton occupancies for select COM wavevectors. Shown on top (a) are the difference between the A exciton occupancies associated with the $K$ and $K'$ valleys (black) as well as their individual contributions (purple and yellow, respectively). Also shown is the total integrated exciton occupancy across nonzero wavevectors (red). Shown below (b) is the exciton occupancy as a function of the wavevector magnitude. Applied parameters include $T = 77~\mathrm{K}$, $N_\mathrm{k}=75$, $R_\mathrm{k}=0.2$, and $R_\mathrm{x}=0.1$ (in units of $2\pi/a$).}
    \label{fig:FIG_X_POPS}
\end{figure}

\begin{figure*}
    \centering
    \includegraphics{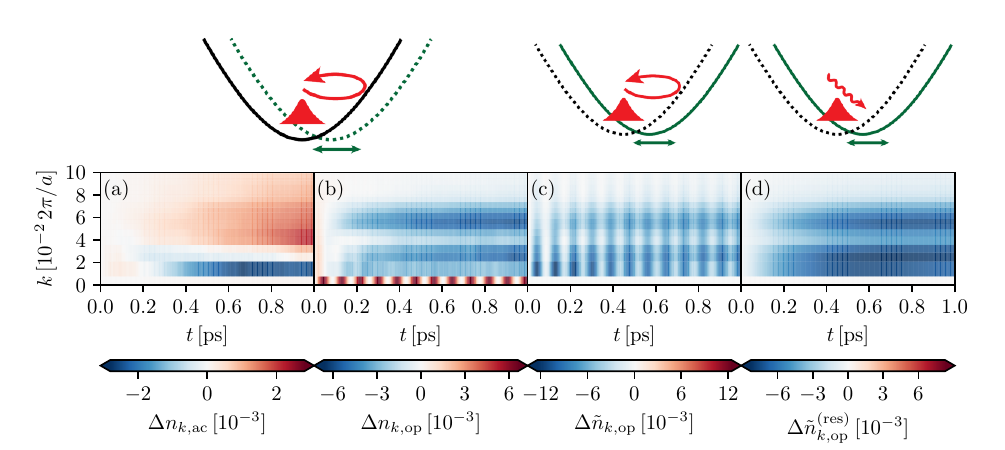}
    \caption{Calculated transient changes in phonon occupancies as a function of wavevector magnitude. Shown are results for the acoustic (a) and the optical (b-d) branches. For the latter, transient occupancies are shown with respect to the unshifted potential (b), the shifted potential (c), and the shifted potential after subtraction of the oscillating component (d), as schematically illustrated on top (see text for details). Applied parameters as in Fig.~\ref{fig:FIG_X_POPS}.}
    \label{fig:FIG_PHOCCS}
\end{figure*}

Having reached qualitative, and at times semi-quantitative, agreement with experimental measurements, our QC framework presents itself as a valuable tool to unveil mechanistic principles driving the fast and dominant component of valley depolarization in MoS$_2$. First, there is an obvious interest in dissecting the excitonic states involved in the dynamics. Combining such a dissection with an extrapolation approach, as done for $\gamma_\infty$ and $\tau_\infty$, is costly and impractical. For that reason, in the following we resort to calculations conducted at one given set of convergence parameters, including $N_\mathrm{k}=75$, $R_\mathrm{k}=0.2$, and $R_\mathrm{x}=0.1$ (in units of $2\pi/a$). Unless noted otherwise, we will also adopt a temperature of 77~K.

Results of the excitonic state dissection are shown in Fig.~\ref{fig:FIG_X_POPS} (a) where, in addition to $\rho_{\mathrm{A},K}-\rho_{\mathrm{A},K'}$, we plot the individual occupancies of the two valley-localized A excitons, $\rho_{\mathrm{A},K}$ and $\rho_{\mathrm{A},K'}$. Both individual occupancies are seen to equilibrate, consistent with a near-complete decay of their difference, which in turn implies a vanishing valley polarization. Also shown in Fig.~\ref{fig:FIG_X_POPS} (a) is the total integrated occupancy of excitons with finite COM wavevector, $\sum_{k\neq0}\rho_{k}$, which is seen to rapidly attain a sizable value at the expense of both $\rho_{\mathrm{A},K}$ and $\rho_{\mathrm{A},K'}$. Such excitons are optically dark, and do not contribute to the TRFR measurements. To further unravel the dynamics among optically-dark excitons, we resolve in Fig.~\ref{fig:FIG_X_POPS} (b) the exciton occupancy as a function of the COM wavevector magnitude (but excluding the overwhelmingly-large contributions from the zero-wavevector A excitons). Here, the exciton occupancy is seen to remain biased towards small wavevectors.

Since our QC framework retains microscopic detail of the phonons, the phonon occupancies can be resolved to the same level of detail as the excitons. In order to elucidate the impact of phonons on valley depolarization, we proceed to specifically consider transient changes in the phonon occupancies, which are obtained as
\begin{align}
    \Delta n_{\bm{k},\mu} = \left\vert z_{\bm{k},\mu}\right\vert^{2} - \left\vert {z_{\bm{k},\mu}}^{(0)}\right\vert^{2},
\end{align}
with ${z_{\bm{k},\mu}}^{(0)}$ as the initial phonon coordinate. Shown in Fig.~\ref{fig:FIG_PHOCCS} (a) and (b) are transient phonon occupancy changes for the optical and acoustic branches, respectively, at finite wavevectors. Considering the acoustic phonons, small-wavevector modes are seen to engage, but only slowly, and past the time scale at which valley depolarization ensues (cf.~Fig.~\ref{fig:FIG_X_POPS}). This can be understood based on the low energies associated with small-wavevector acoustic phonons. The negative change in occupancy indicates that these modes donate energy to the exciton.

The optical phonons, on the other hand, engage comparatively rapidly, and within the relevant time scale of valley depolarization. Their transient occupancy change shows rich dynamics, with a coherent beating contributing prominently in the small-wavevector limit. This beating is indicative of nuclear oscillations induced by a sudden shift in the equilibrium coordinate of the corresponding nuclear potential, induced by transitions of the electronic quantum state. A familiar example of such a sudden potential shift is the one arising upon a vertical electronic excitation within the Franck--Condon model \cite{christensson2012origin}. Such dynamics are captured by our QC framework owing to its non-Markovian nature.

To better understand the transient changes in phonon occupancies, it proves helpful to scrutinize the exciton--phonon interaction Hamiltonian given in Eq.~\ref{eq:H_x_ph}. This Hamiltonian can be reformulated by separating the zero-wavevector phonons from those at finite wavevector, yielding
\begin{align}
    \hat{H}_{\mathrm{x-ph}} =& \sum_{n,\mu}G_{n,n,\mu} \hat{C}^{\dagger}_n \hat{C}^{\phantom{\dagger}}_n \left(z^{*}_{0,\mu} + z_{0,\mu}\right) \nonumber \\& + \sum_{n,m\neq n,\mu}G_{n,m,\mu} \hat{C}^{\dagger}_n \hat{C}^{\phantom{\dagger}}_m \left(z^{*}_{-\bm{k}_{nm},\mu} + z_{\bm{k}_{nm},\mu}\right).
\end{align}
While the zero-wavevector phonons described in the first term are subject to a Franck--Condon interaction, those at finite wavevector only experience a potential shift once exciton coherences are engaged, owing to the offdiagonal excitonic operator $\hat{C}^{\dagger}_n \hat{C}^{\phantom{\dagger}}_m$. In our simulations, we initialize the quantum state as the A exciton associated with the $K$ valley, which represents a (diagonal) population term for which the potentials of finite-wavevector phonons remain centered at the original minimum. Once exciton dynamics ensues, however, we can expect the potentials to transiently shift due to the rise in exciton coherences. Notably, this shift will induce a superficial inflation of phonon occupancies, so long as these occupancies are evaluated with respect to the unshifted potential. This is remedied by instead evaluating occupancies with respect to the instantaneous potential minimum, by considering
\begin{equation}
    \Delta \tilde{n}_{\bm{k},\mu} = \left\vert z_{\bm{k},\mu} - f_{\bm{k},\mu}\right\vert^{2} - \left\vert {z_{\bm{k},\mu}}^{(0)} - f_{\bm{k},\mu}\right\vert^{2}.
\end{equation}
The shift is obtained straightforwardly through
\begin{equation}
    f_{\bm{k},\mu}=-\frac{\braket{a|\partial z^{*}_{\bm{k},\mu}\hat{H}_{\mathrm{x-ph}}(\bm{z})|a}}{\omega_{\bm{k},\mu}}.
\end{equation}
Here, $a$ is the active surface, being representative of the instantaneous quantum state within the FSSH formalism \cite{tully1990molecular}.

\begin{figure}
    \centering
    \includegraphics{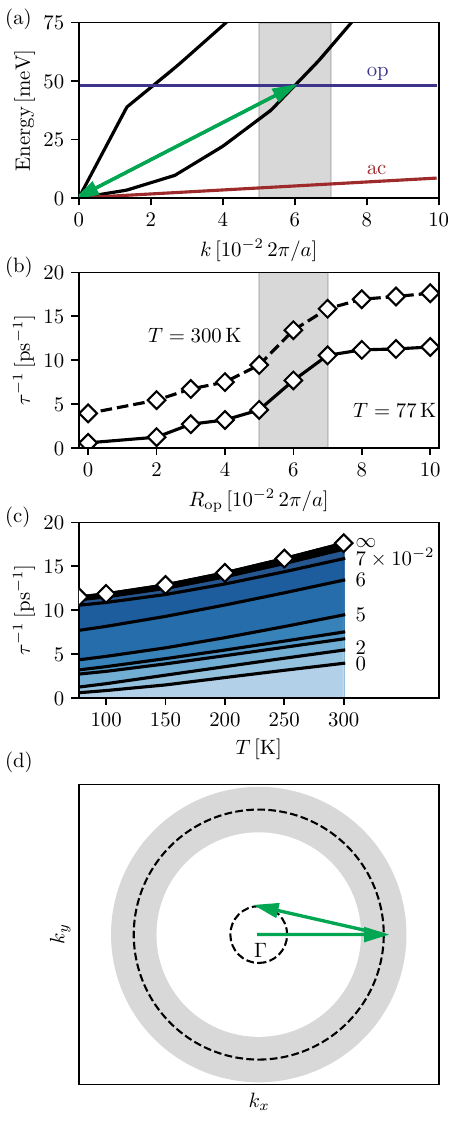}
    \caption{Exciton and phonon dispersions (a). Shown are eigenenergies of the excitonic Hamiltonian operator, $E_n$ (black), as well as the energies of the optical phonon, $\omega_0$ (purple), and the acoustic phonon (red). Shaded area indicates the resonance between the lowest-energy exciton band and the optical phonon branch. Shown below are the valley depolarization rates as a function of a truncation radius imposed on the optical phonons, $R_\mathrm{op}$ (b), at temperatures of 77 K and 300 K. Full temperature resolved results are also included (c). Here, curves follow increments of $10^{-2}~2\pi/a$, with select multiples printed on the right. Applied parameters as in Fig.~\ref{fig:FIG_X_POPS}. Relevant exciton--phonon scattering pathway is shown schematically at the bottom (d).}
    \label{fig:FIG_TRUNC}
\end{figure}

Values of $\Delta \tilde{n}_{\bm{k},\mu}$ for the optical phonons are presented in Fig.~\ref{fig:FIG_PHOCCS} (c). These transient phonon occupancy changes follow a more comprehensible pattern, with a single pronounced beating manifested throughout all wavevectors. Subtracting this beating (by a fitting to a sine wave) yields the residual traces shown in Fig.~\ref{fig:FIG_PHOCCS} (d), which inform on energy transfer between the optical phonon branch and the exciton. Here, two dominant features are apparent, with the first being concentrated at small wavevectors, much like the acoustic phonon, and the second being centered at a wavevector $k=0.06$ in units of $2\pi/a$ (with $a$ the lattice constant). This wavevector value happens to coincide with the crossing of the lowest exciton band and the optical phonon, as shown in Fig.~\ref{fig:FIG_TRUNC} (a). The crossing signifies that an excitonic transition from the $\Gamma$ point to $k=0.06$ may ensue through the absorption of an optical phonon, which conserves both total energy and wavevector, and which we refer to as exciton--phonon resonance. While such process is analogous to vibronic resonance in molecular systems \cite{womick2011vibronic, christensson2012origin, tiwari2013electronic, fuller2014vibronic, romero2014quantum, palecek2017quantum, thyrhaug2018identification, bakulin2016real, fujihashi2017effect, morrison2017evidence, tempelaar2018vibronic, unger2022modulating}, it distinguishes itself by its explicit wavevector conservation requirement. This results from wavevector being a good quantum number in crystalline solids, in contrast to typical molecular systems where relevant excitations are spatially localized and thus have zero momentum.

To more directly assess the impact of exciton--phonon resonance on the valley polarization dynamics, we present complementary simulations in which we  introduce a truncation radius for the optical phonons around the $\Gamma$ point, denoted $R_\mathrm{op}$. Shown in Fig.~\ref{fig:FIG_TRUNC} (b) is the valley depolarization rate, $\tau^{-1}$, calculated at 77 K as a function of $R_\mathrm{op}$. Here, a reduction in $R_\mathrm{op}$ is seen to yield a near-monotonic decrease in $\tau^{-1}$, but with much of the decrease occurring specifically around $k=0.06$. This substantiates that the resonance between the lowest exciton band and the optical phonon forms a dominant driver of valley depolarization. We find the resonance effect to be operative across the full range of temperatures, as evident from comparative results at 300~K shown in Fig.~\ref{fig:FIG_TRUNC} (b), as well as an extended temperature-resolved analysis presented in Fig.~\ref{fig:FIG_TRUNC} (c). These results also underscore that acoustic phonons by themselves provide only a minor contribution to valley depolarization, gauging from the comparatively small values of $\tau^{-1}$ obtained for $R_\mathrm{op} = 0$, where optical phonons are omitted entirely.

It may seem surprising that the exciton--phonon resonance does not lead to an appreciable accumulation of exciton occupancy at $k=0.06$ (cf.~Fig.~\ref{fig:FIG_X_POPS}). Indeed, optical phonons at $k=0.06$ should initially facilitate a scattering from an A exciton state with vanishing COM wavevector towards an exciton state at $k=0.06$. It bears noting, however, that the same phonons promote a scattering from a $k=0.06$ exciton back to the $\Gamma$ point region. Moreover, due to the two-dimensional nature of the Brillouin zone, this back scattering may terminate at a wavevector close to, but not coincident with, the $\Gamma$ point, while lattice momentum is still conserved. This is schematically depicted in Fig.~\ref{fig:FIG_TRUNC} (d). Such back-and-forth scattering behavior not only explains the small exciton occupancy at $k=0.06$, but also is consistent with the rapid accumulation of exciton occupancy at small but finite wavevectors apparent in Fig.~\ref{fig:FIG_X_POPS} (b).
 
\section{Conclusion}

In summary, we have applied a QC simulation framework in order to investigate the fast and dominant valley depolarization component in monolayer MoS$_2$. The framework was recently developed based on a mean-field formalism \cite{krotz2024mixed}, and is here extended by adopting FSSH. It allows exciton--phonon interactions to be incorporated at a microscopic, non-perturbative, and non-Markovian level, while keeping the computational cost sufficiently low for it to enable dynamical calculations at material sizes approaching the thermodynamic limit. Resulting simulations yielded optical linewidths and valley depolarization times consistent with experimental measurements \cite{dey2016OpticalCoherenceAtomicMonolayer, dalconte2015ultrafast} across temperatures. The simulations furthermore retained explicit information on transient phonon occupancies, the analysis of which showed a resonance between the dominant optical phonon branch and the lowest-energy exciton band to present a major pathway for valley depolarization. Such exciton--phonon resonance presents itself as a solid-state analog of vibronic resonances in molecular systems \cite{womick2011vibronic, christensson2012origin, tiwari2013electronic, fuller2014vibronic, romero2014quantum, palecek2017quantum, thyrhaug2018identification, bakulin2016real, fujihashi2017effect, morrison2017evidence, tempelaar2018vibronic, unger2022modulating}, although it differs by its wavevector-resolved selection rules due to wavevector being a good quantum number in crystals. It will be interesting to explore the significance of similar resonance effects in a wider range of materials, for which non-perturbative and/or non-Markovian simulations may prove essential.

Our simulations showed the importance of optical phonons in driving valley depolarization in MoS$_2$, while acoustic phonons were found to only provide a minor contribution to this process, owing to their comparatively-long characteristic time scale. Our results thus suggest the suppression of optical phonons to present a prime target for engineering efforts seeking to extend valley depolarization times. They also suggest that exciton--phonon resonances may be leveraged to create a transient population of polarized dark excitons. Such an approach might be realized through band engineering which enables an exciton--phonon resonance to populate the conduction band $\Lambda$ valley wherein exchange-driven depolarization may be suppressed.

Moving forward, it will be of interest to expand our analysis to other members of the monolayer transition-metal dichalcogenide family. While in the present study we resorted to \emph{ab initio} parametrized models for the quasiparticle band structure, electron--hole interactions, and the carrier--phonon interactions, we should reiterate that the applied QC framework is not fundamentally reliant on such models, and can be directly interfaced with \emph{ab initio} calculations. It will be particularly worthwhile to adopt \emph{ab initio} band structures, which would realistically capture electron transitions towards the $\Lambda$ region of the conduction band, which is predicted to govern rapid valley polarization dynamics in WSe$_2$ \cite{selig2020suppresion}. Such \emph{ab initio} extensions would also allow the slow valley depolarization component of MoS$_2$ to be studied, thereby expanding upon the analysis of the fast component presented in this Article. Such advancements of our framework not only present a path towards a holistic investigation of valley depolarization across monolayer transition-metal dichalcogenides, but also establish a valuable modeling capability for studying strong carrier--phonon interactions in a broad range of materials.

\section{Acknowledgements}

The authors thank Denis Karaiskaj for providing the measured optical linewidth data shown in Fig.~\ref{fig:FIG_LW_TRFR} and Stefano Dal Conte for providing the measured TRFR data, also shown in Fig.~\ref{fig:FIG_LW_TRFR}. This material is based upon work supported by the National Science Foundation under Grant No.~2145433.

\bibliography{bibliography}

\end{document}


\title{Supporting Information for ``Surface Hopping Simulations Show Valley Depolarization Driven by Exciton--Phonon Resonance''}
\author{Alex Krotz}
\author{Roel Tempelaar}
\email{roel.tempelaar@northwestern.edu}

\maketitle

\section{Determination of calculated optical linewidth}\label{sec:linewidth}

\begin{figure}
    \centering
    \includegraphics{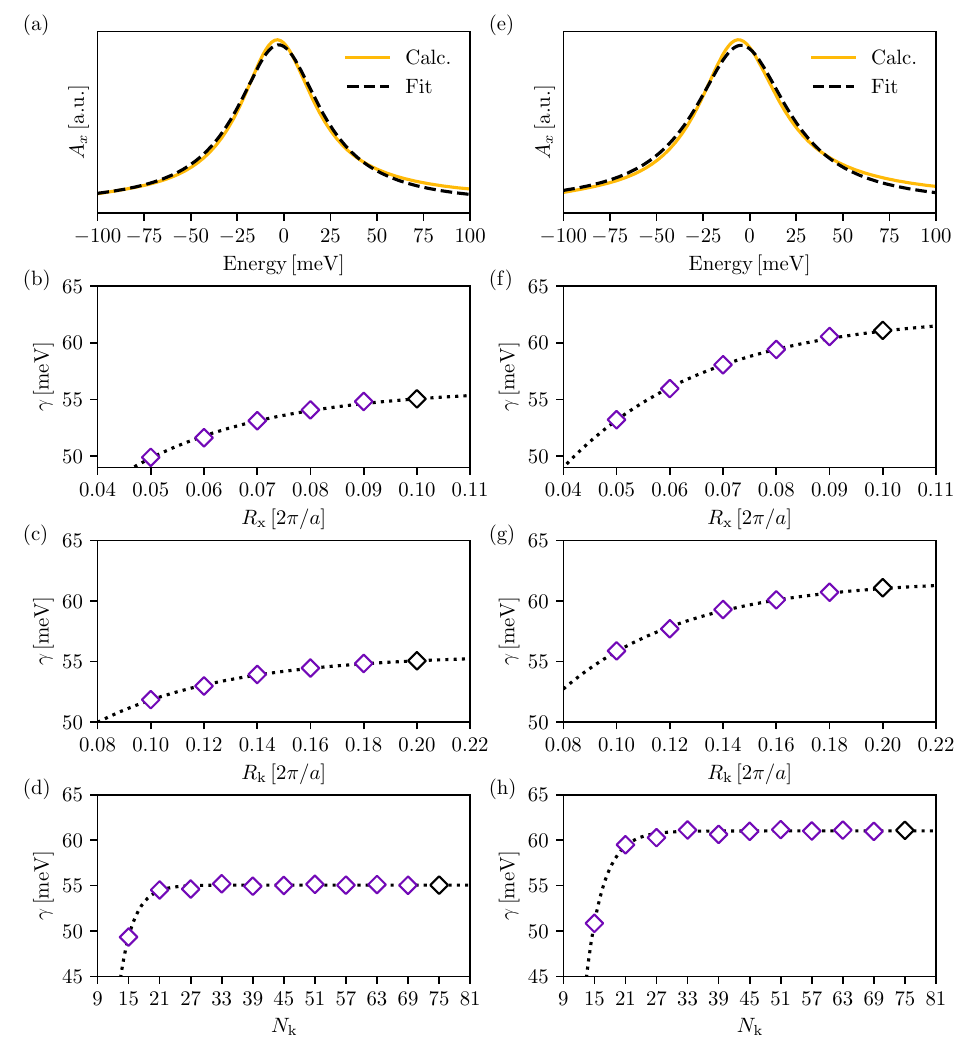}
    \caption{Absorption spectra (yellow curve), calculated at $T=77~\mathrm{K}$ (a) and $T=300~\mathrm{K}$ (e) and using truncation parameters $N_\mathrm{k}=75$, $R_\mathrm{k}=0.2$, and $R_\mathrm{x}=0.1$ (in units of $2\pi/a$), shown alongside a Lorentzian fit used to determine the optical linewidth (black dash). Shown below are calculated linewidths (diamonds) at varying values of the truncation parameters $T=77~\mathrm{K}$ (b-d) and $T=300~\mathrm{K}$ (f-h). Also shown are exponential fits used to determine the asymptotic linewidths, $\gamma_\infty$ (dotted curve).}
    \label{fig:lw_conv}
\end{figure}

The calculated absorption spectrum associated with polarization $\lambda$ is obtained by means of the Fourier transform \cite{berne1970CalculationTimeCorrelation},
\begin{equation}
    A_{\lambda}(\omega) =\Re\left[ \int_{0}^{\infty}\dd\tau\, e^{i\omega\tau}R_{\lambda}(\tau) \right],\label{eq:abs_spec}
\end{equation}
with $\Re$ denoting the real part and with the dipole correlation function given by
\begin{equation}
    R_{\lambda}(\tau)=\langle 0\vert \hat{P}_{\lambda}\vert\Psi(\tau)\rangle.
\end{equation}
Here, $\hat{P}_{\lambda}$ denotes the electronic momentum operator, a discussion of which is presented in Ref.~\citenum{krotz2024mixed}. Further, $\ket{0}$ denotes the electronic ground state, with completely filled valence bands and empty conduction bands (referred to as the Fermi vacuum), and $\ket{\Psi(\tau)}$ represents the electronic wavefunction at time $\tau$. The electronic wavefunction is initialized as $\ket{\Psi(0)} = \hat{P}^{\dagger}_{\lambda} \ket{0}$. By resorting to the electronic wavefunction, rather than active surfaces, we follow a conventional approach to evaluating optical response functions \cite{tempelaar2013surface, landry2013correct}, although noting that evaluations based on active surfaces within fewest-switches surface hopping (FSSH) have been proposed \cite{tempelaar2017generalization, bondarenko2023overcoming}. The absorption spectrum is then convolved with a Lorentzian in order to account for non-phononic contributions to the optical linewidth \cite{selig2016ExcitonicLinewidthCoherence, chan2023exciton} (these include radiative electron--hole recombination and inhomogeneous broadening). In doing so, a temperature-independent full width at half maximum (FWHM) of 40 meV is applied, which is based on a fit to the experimental absorption spectrum measured at 5 K \cite{dey2016OpticalCoherenceAtomicMonolayer}. The principal peak apparent in the absorption spectrum is then fitted to a Lorentzian, the FWHM of which yields the optical linewidth. Calculated linewidths and corresponding fits are shown in Fig.~\ref{fig:lw_conv} (a) and (e) at temperatures $T=77~\mathrm{K}$ and $T=300~\mathrm{K}$, respectively.

\section{Breakdown of Optical Linewidth}

\begin{figure}
    \centering
    \includegraphics[width=0.5\linewidth]{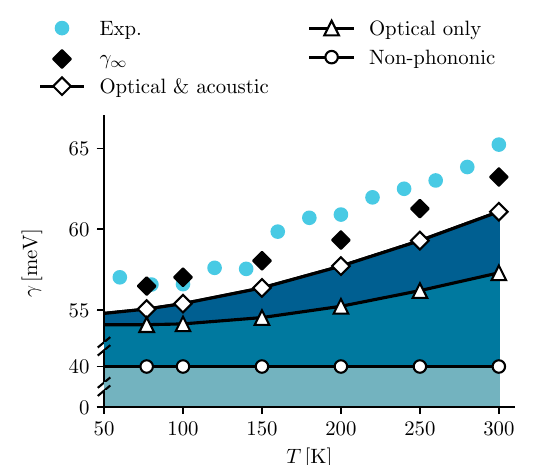}
    \caption{Breakdown of the optical linewidth into a non-phononic contribution (open circles), as well as contributions due to the optical (open triangles) and acoustic (open diamonds) phonons. Also shown are experimental results (cyan circles) and calculated linewidths in the asymptotic limit of the convergence parameters (closed diamonds).}
    \label{fig:lw_delta}
\end{figure}

In Fig.~\ref{fig:lw_delta}, we present a breakdown of the optical linewidth into a non-phononic contribution as well as contributions due to optical and acoustic phonons. The temperature-independent non-phononic contribution amounts to 40~meV, as detailed in Sec.~\ref{sec:linewidth}. A comparative calculation including only optical phonons allows us to assess the contribution of such phonons, whereupon the contribution due to acoustic phonons is assessed by a comparative calculation in which both optical and acoustic phonon branches are included. All calculations are performed at convergence parameter values $N_{\mathrm{k}}=75$, $R_{\mathrm{k}}=0.2$, and $R_{\mathrm{X}}=0.1$ (in units of $2\pi/a$). In the low-temperature limit, the temperature dependence of the linewidth originates predominantly from acoustic phonons, until the optical phonons begin contributing for $T>150$~K.

\section{Convergence of Optical Linewidth}\label{sec:conv_linewidth}

In determining the optical linewidth in the asymptotic limit of the convergence parameters, $\gamma_\infty$, we follow the same extrapolation approach as in Ref.~\citenum{krotz2024mixed}. Accordingly, we take as a reference point the convergence parameter values $N_{\mathrm{k}}=75$, $R_{\mathrm{k}}=0.2$, and $R_{\mathrm{X}}=0.1$, and then calculate the linewidth while individually varying each of these values, the result of which is fitted to the product of three exponential functions
\begin{equation}
    \gamma(R_{\mathrm{k}},R_{\mathrm{X}},N_{\mathrm{k}}) = \gamma_{\infty}f_{\mathrm{N}}(N_{k})f_{\mathrm{k}}(R_{\mathrm{k}})f_{\mathrm{X}}(R_{\mathrm{X}}).\label{eq:lw_exp}
\end{equation}
Here, $f_{\mathrm{N}}(N_\mathrm{k}) \equiv 1 - \exp(-b_\mathrm{N} N_\mathrm{k} + c_\mathrm{N})$, with $b_\mathrm{N}$ and $c_\mathrm{N}$ as fit parameters, and similarly for $f_{\mathrm{k}}(R_{\mathrm{k}})$ and $f_{\mathrm{X}}(R_{\mathrm{X}})$. Calculated linewidths and corresponding fits at temperatures $T=77~\mathrm{K}$ and $T=300~\mathrm{K}$ are shown in Fig.~\ref{fig:lw_conv} (b-d) and (f-h), respectively.

\section{Convergence of Valley Depolarization Time}

\begin{figure}
    \centering
    \includegraphics[]{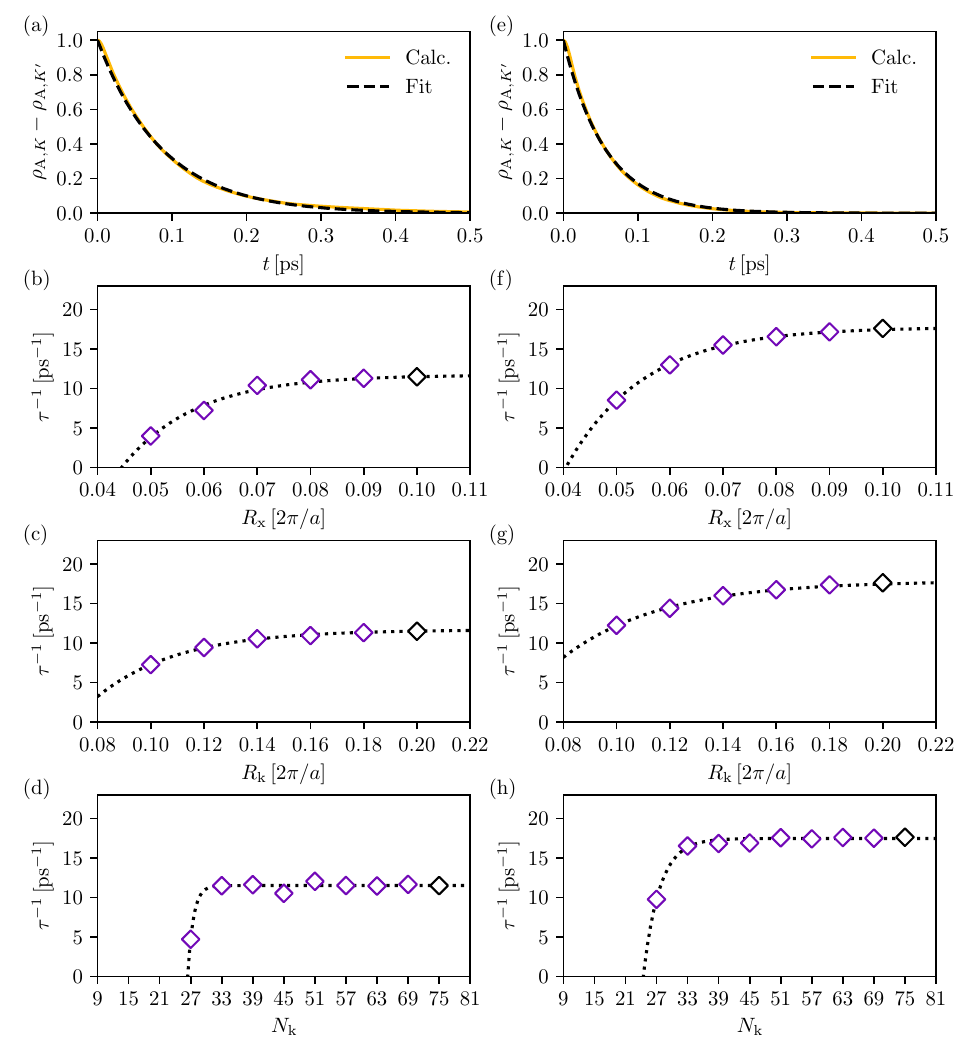}
    \caption{Same as Fig.~\ref{fig:lw_conv} but for the valley depolarization rate. Shown on top are calculated transient differences between the A exciton occupancies in the $K$ and $K'$ valleys (yellow curve) at $T=77~\mathrm{K}$ (a) and $T=300~\mathrm{K}$ (e), alongside the exponential fits (black dash) used to determine the valley depolarization times. Shown below are calculated depolarization rates (diamonds) at varying values of the truncation parameters $T=77~\mathrm{K}$ (b-d) and $T=300~\mathrm{K}$ (f-h). Also shown are exponential fits used to determine the asymptotic rates, $\tau^{-1}_\infty$ (dotted curve).}
    \label{fig:trfr_conv_77}
\end{figure}

In determining the converged valley depolarization time, we first perform an exponential fit to the difference between the A exciton populations in the $K$ and $K'$ valleys, $\rho_{\mathrm{A},K}-\rho_{\mathrm{A},K'}$, using the function
\begin{equation}
    P(t)=\exp\left(-t\; \tau^{-1}\right).
\end{equation}
In doing so, we only consider the time interval where $P(t)>0.01$ (which is determined iteratively). Calculated transient population differences and corresponding fits are shown in Fig.~\ref{fig:trfr_conv_77} (a) and (e) at temperatures $T=77~\mathrm{K}$ and $T=300~\mathrm{K}$, respectively. Similarly to the determination of the converged linewidth, detailed in Sec.~\ref{sec:conv_linewidth}, we vary the values of the convergence parameters, and fit the calculated depolarization times to the product of three exponential functions,
\begin{equation}
    \tau^{-1}(R_{\mathrm{k}},R_{\mathrm{X}},N_{\mathrm{k}}) = \tau_\infty^{-1}f_{\mathrm{k}}(R_{\mathrm{k}})f_{\mathrm{X}}(R_{\mathrm{X}})f_{\mathrm{k}}(N_{\mathrm{k}}),\label{eq:trfr_exp}
\end{equation}
in order to obtain the converged depolarization time, $\tau_\infty$. Calculated depolarization times and corresponding fits are presented in Fig.~\ref{fig:trfr_conv_77} (b-d) and (f-h) at temperatures $T=77~\mathrm{K}$ and $T=300~\mathrm{K}$, respectively.

\bibliography{bibliography}